# DUST FORMATION IN ELECTRIC ARC FURNACE: BIRTH OF THE PARTICLES

Anne-Gwénaëlle Guézennec[1*], Jean-Christophe Huber[2], Fabrice Patisson[1], Philippe Sessiecq[1], Jean-Pierre Birat[2], Denis Ablitzer[1]

1 – *Laboratoire de Science et Génie des Matériaux et de Métallurgie, Parc de Saurupt, 54042 Nancy Cedex, France*
* guezenne@mines.inpl-nancy.fr
2 – *IRSID, Voie Romaine, BP 30320, 57283 Maizières lès Metz, France*

**Abstract.** The characterization of electric arc furnace (EAF) dust shows that bubble burst at the liquid steel surface is the principal source of dust emission. We have therefore developed an experimental device for studying this phenomenon. As in the case of the air-water system, the bubble-burst gives birth to two types of droplets: film drops and jet drops. The jet drop formation is observed with high-speed video. The film drop aerosol is collected on filters and, then, characterized by means of SEM, granulometric and gravimetric analyses. Results are presented and discussed. The quantification of both types of projections leads to the conclusion that the film drop projections represent the major source of dust. The amount of film drops greatly decreases with the parent bubble size. Under 4.5 mm in bubble diameter, no film drops are formed. Decreasing enough the bubble size would therefore represent an effective solution for reducing drastically the EAF dust emission.

## INTRODUCTION

The Electric Arc Furnace (EAF), designed for steelmaking from recycled scrap iron (see figure 1), also co-produces between 15 to 25 kg of dust per ton of steel. Dust formation is strongly linked to the process which can be divided into five steps:
- furnace charging: the scrap and the additives (lime, coal…) are loaded into special charging buckets which are then emptied into the furnace;
- melting: an electric arc is created between the graphite electrodes and the scrap which entails the charge melting and the formation of a steel bath covered by a slag layer, volatile solute species (e.g. zinc) begin to be removed;
- refining: in this step of the process, phosphorus is removed from the steel bath by interfacial reactions between the slag and the liquid metal, injection of oxygen promotes the decarburization reaction with dissolved carbon and bubbles of carbon monoxide (CO) are formed, which helps to remove other dissolved gases;
- slag foaming: the CO-bubbles crossing the slag layer make it foam, the foaming process being enhanced by the addition of coal powder;
- casting: after the composition and the temperature of the bath have been controlled, the liquid steel is cast.

During the process, the fumes are extracted through an aperture in the furnace roof. These are post-combusted, cooled, and cleaned from the transported dust, which is collected in large bag filters. This dust contains hazardous, leachable elements such as zinc, lead or cadmium which require EAF dust to be stored in specific landfills.

In order to propose economically feasible solutions for both recycling and/or reducing EAF dust, the understanding of the dust formation is necessary. The present paper describes the different mechanisms of formation identified thanks to a morphological and mineralogical characterization of various dust samples, and then focuses on the study of the main source of emission, i.e. bubble burst at the surface of the liquid bath. An original experimental device was designed in order to understand and quantify precisely this phenomenon. The results of the experimental study are presented and discussed.

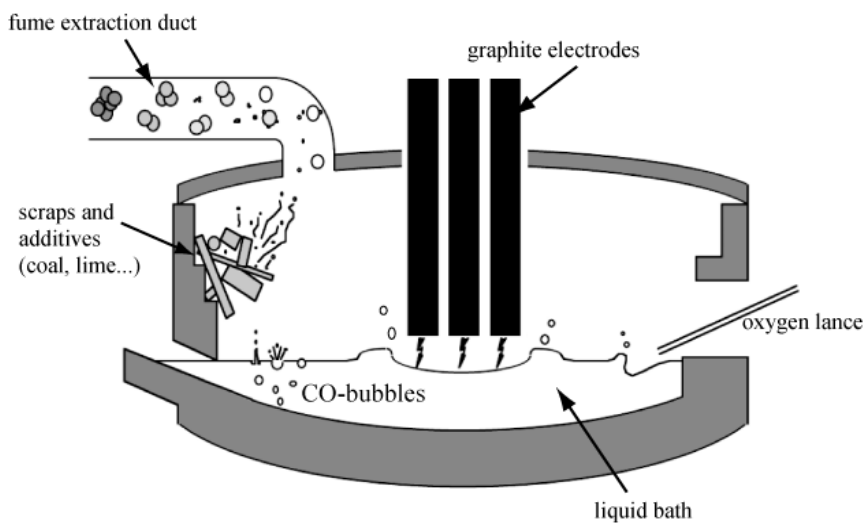

*Figure 1. Schematic representation of an Electric Arc Furnace*

## CHARACTERIZATION OF ELECTRIC ARC FURNACE DUST

The investigations regarding the morphology and mineralogy of the particles contained in EAF dust give useful information for the identification of the dust formation mechanisms. Several dust samples coming from different industrial furnaces were observed by SEM (Scanning Electron Microscopy) and analyzed by EDS (Energy Dispersive Spectrometry). As shown in figure 2, EAF dust particles cover a wide range of sizes. To simplify the survey of the morphologies, we distinguished two categories of particles: large particles from a few dozen to a few thousand micrometres, and finer particles lower than 20 μm.

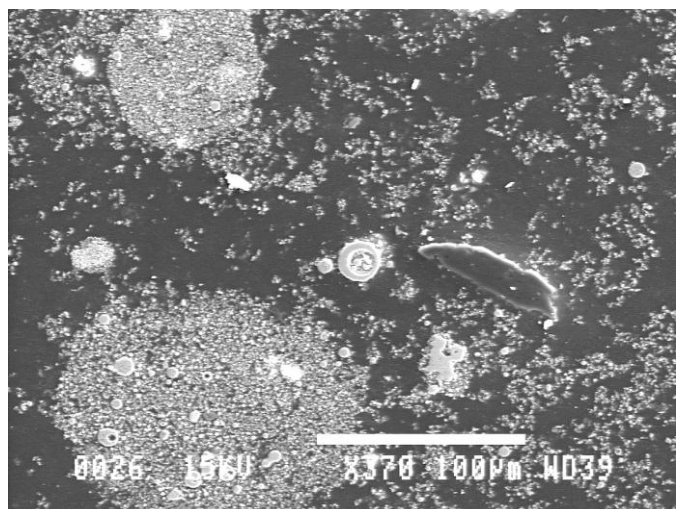

*Figure 2. EAF dust*

### 1. Large particles

Three morphological types belong to this category. The first one is composed of particles of coal and lime (figures 3 and 4). Their sizes vary between 20 and 500 μm and their shapes are irregular. This morphology indicates they come from the direct fly-off of solid particles during the introduction of powder materials into the EAF (scrap, coal for slag foaming, additions, recycled dust, etc.).

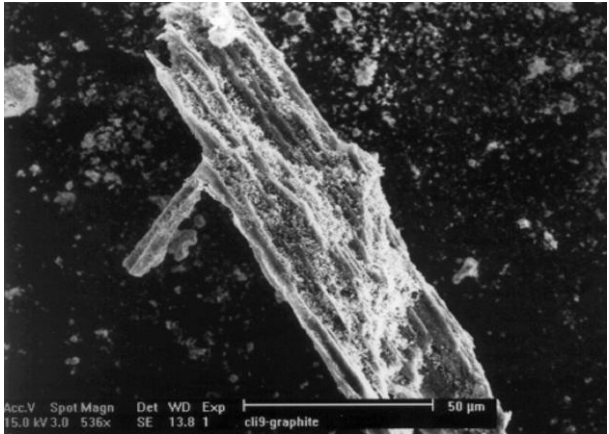

*Figure 3. Coal particle*

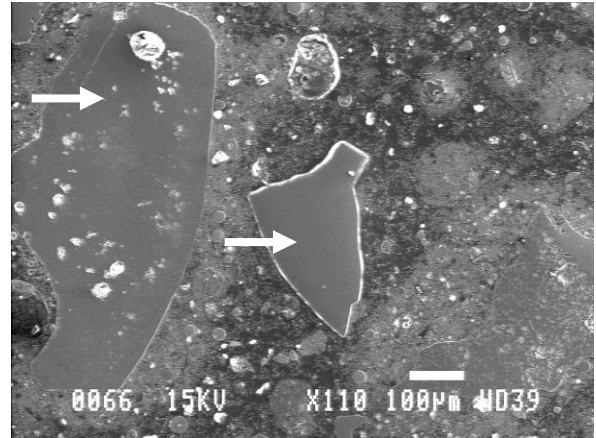

*Figure 4. Lime particle*

The second one is made up of sphere-like particles whose sizes range from 20 to 200 µm (figure 5). Their chemical composition corresponds to that of the slag (Ca, Al, Fe, Si…). They probably result from a phenomenon of liquid droplets projection at the impact points of the arc or of the oxygen jet on the liquid bath.

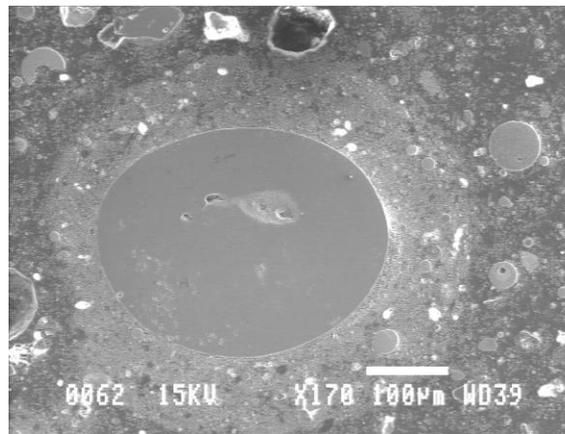

*Figure 5. Sphere-like particle whose composition corresponds to that of the slag*

The third morphological type corresponds to agglomerates of fine particles (figure 6) similar to those presented in the following part. Their sizes vary between 20 and 1000 µm. They are fragile and break up easily. Thus, these particles are likely formed by low-temperature agglomeration (e.g. in filters).

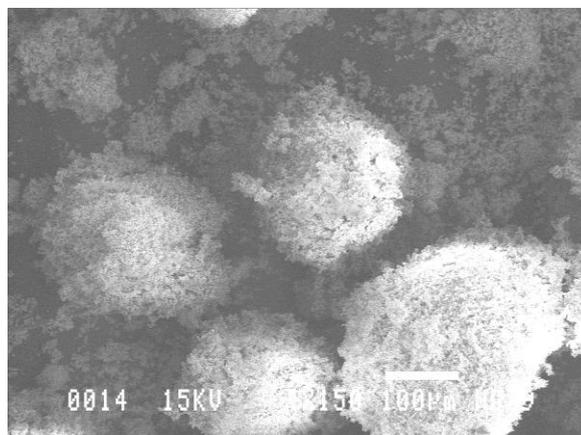

*Figure 6. Large agglomerates of fine particles*

Contrary to the agglomerates of fine particles, the first two types of particles are hardly present or even completely absent from the dust samples. Their presence suggests an excessive fume extraction flow rate or a bad control of the melting and addition processes.

## 2. Fine particles

Fine particles, whose sizes are below 20 μm, account for the major part of EAF dust. A small proportion of those particles corresponds to monocristals of zinc oxide (figure 7). They are easily identifiable thanks to their facetted aspect. Their size rarely exceeds a few hundred nanometres.

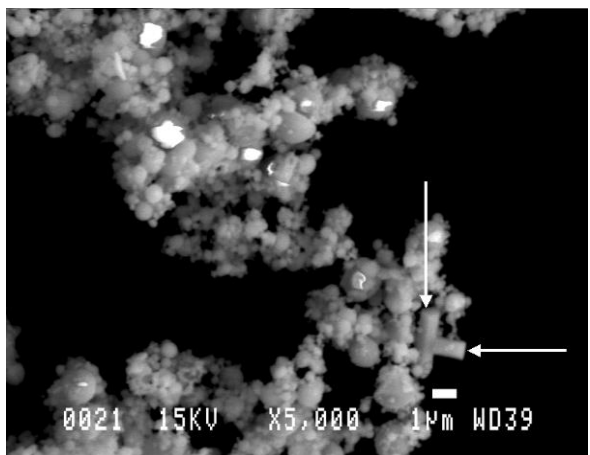

*Figure 7. Zincite monocristals*

The other particles are spherical. Their size varies from 0.2 to 20 μm. We found three types of spheres which differ from each other because of their mineralogy:
- homogeneous spheres whose composition corresponds either to the slag or to the steel bath with an enrichment in zinc; they are often hollow when they are larger than 2 or 3 μm (figure 8);
- heterogeneous spheres made up of a slag phase and a steel phase enriched in zinc (figure 9); some of them display an iron-rich dendritic structure buried inside a vitreous phase;
- submicronic spheres of pure zincite.

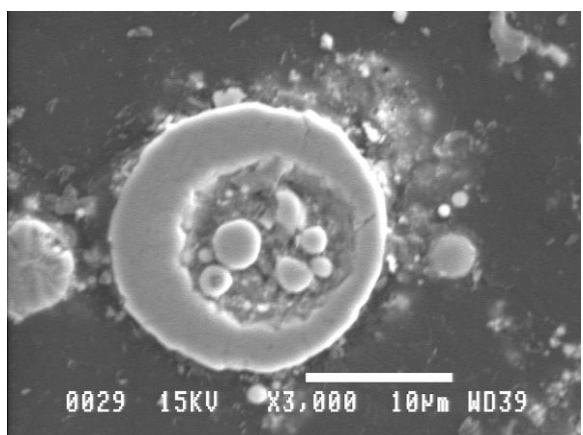

*Figure 8. Several full spheres and one hollow sphere*

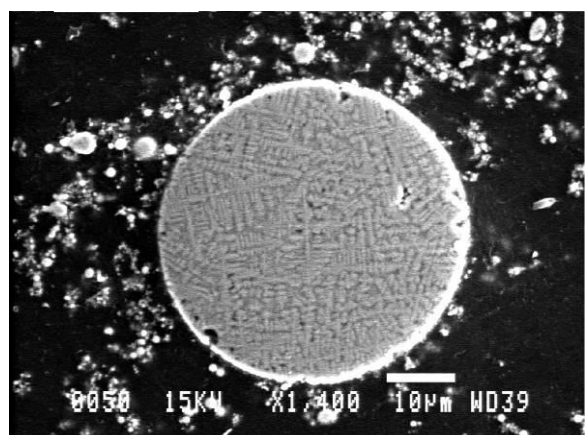

*Figure 9. Heterogeneous sphere*

The zincite spheres, like the monocristals, form during the condensation of the vapors of zinc contained in the EAF fumes [1, 2]. The other spheres represent the major part of the dust observed. They come from the projection of liquid droplets. On account of their sizes, they are thought to be emitted by the burst of CO-bubbles coming from the decarburization of the steel bath [1, 3, 4, 5].

The finest particles, whose sizes are lower than 2 or 3 µm, are frequently agglomerated to each other or around a bigger particle (figure 10). The sizes of these agglomerates vary from 5 to 20 µm; a few of them reach 50 µm and some show signs of partial sintering also noted by Cruells *et al.* [6]. In this case, the agglomeration took place inside the furnace or the fume extraction ducts at high temperature.

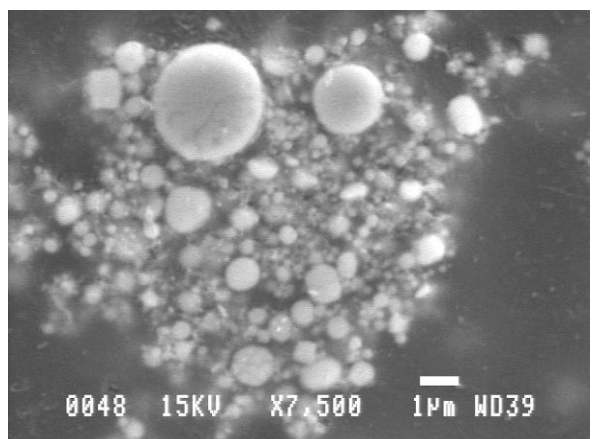

*Figure 10. Agglomerates of fine particles*

### I.3. Interpretation

The dust collected in bag filters at the end of the EAF fume extraction system is the final product of a series of phenomena, such as the emission of particles from the steel bath, the transport of these particles by the gas flow in the fume extraction system, the in-flight physico-chemical transformations they undergo, etc. The results of the morphological analysis of the EAF dust show that the dust formation process takes place in two steps: first, the emission of dust "precursors", i.e. vapors, metal droplets, and solid particles, inside the furnace; second, the conversion of those precursors into dust by agglomeration and physico-chemical transformations.

From the different types of particles displayed previously, five emission mechanisms of dust precursors have been identified (see figure 11):
- volatilization, especially localized at the hot spots in the arc zone (1) and the oxygen jet zone (1'), but taking place as well in the CO bubbles;
- projection of droplets at the impact points of the arc (2) and of the oxygen jet (2') on the steel bath;
- projection of fine droplets by bursting of CO bubbles (3) coming from the decarburization of the steel bath;
- bursting of droplets (4) in contact with an oxidizing atmosphere within the surface; the occurrence of this phenomenon, which can be classed as a bubble-burst mechanism, is uncertain in EAF;
- direct fly-off of solid particles (5) during the introduction of powder materials into the EAF (scrap, coal for slag foaming, additions, recycled dust, etc.).

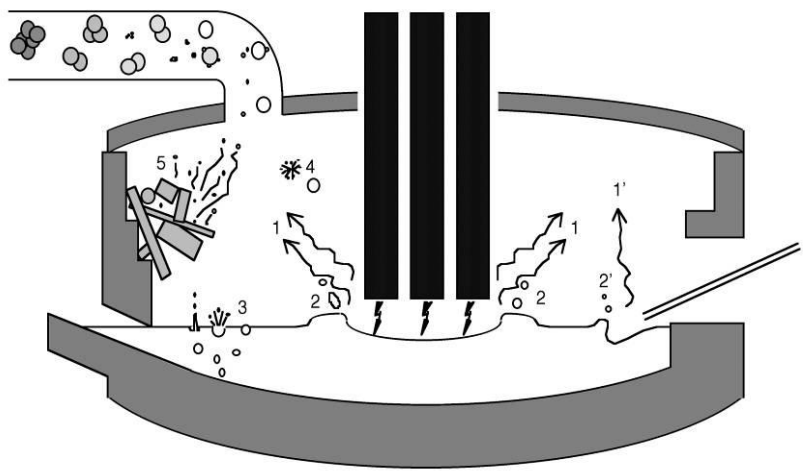

*Figure 11. Schematic representation of the mechanisms of dust emission in EAF*

According to the preceding analysis and experimental quantifications of each mechanism made by Birat *et al.* [7], the prevailing mechanisms of dust precursor emission appear to be the volatilization (27 % of the dust) and the bursting of CO bubbles (60 % of the dust). The direct fly-off of solid particles remains very limited if sufficient operating cautions are taken. As for the projections at the impact points of the arc or of the oxygen jet, most of them fail to be carried up by the fume extraction system, due to their size, and fall back into the liquid bath.

The precursors are further transformed during their transport within the furnace and then in the fume extraction system. They can undergo physical transformations: condensation of the vapors, rapid solidification of the fine projections in contact with a colder atmosphere, in-flight agglomeration and coalescence of dust particles. The precursors can also be modified by chemical reactions (e.g. oxidation) with the carrier gas, whose temperature and composition vary, and, they can possibly react with other precursor particles. For a reaction between condensed phases (liquid or solid) to occur, particles must first be brought into contact. Therefore, there is a strong link between the mechanisms of agglomeration and the chemical evolution [8].

**DUST FORMATION BY BUBBLE BURSTING**

**1. Theory**
The projection of liquid steel and slag droplets by bursting of CO bubbles has been recognized as the principal mechanism of dust emission in EAF. Very few studies about bubble-burst at the surface of liquid metal have been reported [9]. However, in order to understand the phenomenon, useful results and observations can be found in the abundant literature about the air-water system. From these studies, the bubble-burst process can be split up into three steps which give rise to two types of droplets (figure 12).

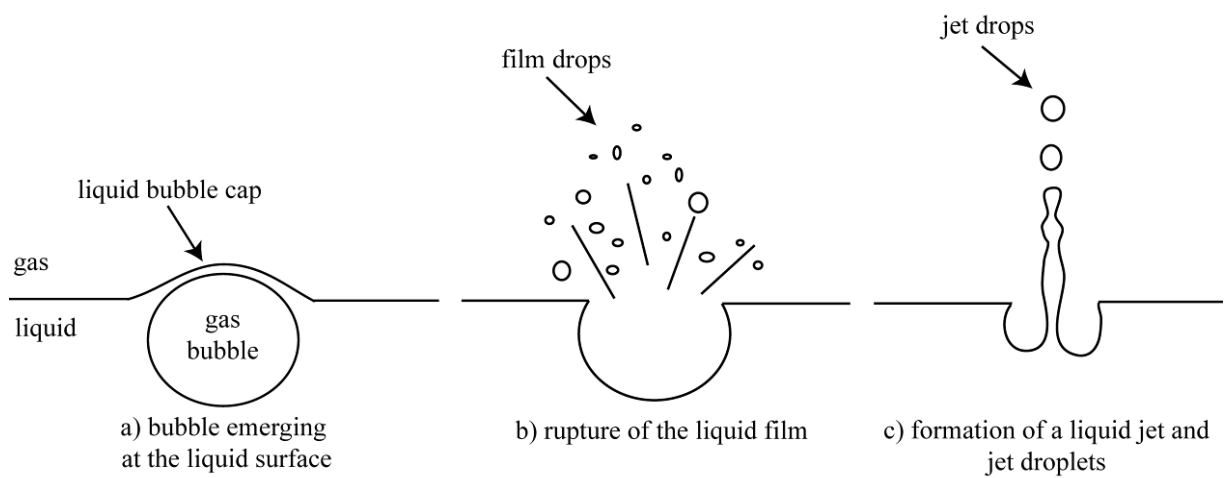

*Figure 12. Schematic representation of the burst of a bubble on a liquid surface*

When emerging at the surface (figure 12a), a bubble lifts up a liquid film that progressively gets thinner under the influence of drainage, when the bubble comes to rest. The shape of a bubble floating at the surface of a liquid can be determined by following the approach proposed by Unger *et al.* [10].

As the film reaches a critical thickness, it breaks up and the bubble cap is disintegrated into fine droplets called film drops (figure 12b). Many authors [11, 12, 13, 14] studied the number and size of film drops as a function of the bubble size. The number is proportional to the surface of the film. The size distribution is wide: from 0.3 to 500 μm.

After the disruption of the bubble cap, the cavity remaining at the liquid surface closes up, creating an upward Rayleigh jet that is unstable and can break up into droplets usually called jet drops (figure 12c). The number of jet drops never exceeds ten and decreases when the bubble size increases [11, 14]. Their sizes have been found to range between 0.1 and 0.18 times the diameter of their parent bubble for air-water system [15, 16].

## 2. Experimental apparatus

In order to study dust emission from bubble burst in liquid steel, we set up an original experimental device (figure 13) using a vacuum induction melting furnace (Leybold) modified in order to operate at atmospheric pressure under an argon atmosphere. The aims of the experiments are to clarify the way the bubbles burst at a liquid steel bath surface and to quantify the resulting emissions, i.e. film drops and jet drops.

The steel charge (750 g of a commercial steel grade XC38) is melted in an alumina crucible (45 mm inside diameter, 70 mm height), fitted in a graphite susceptor. This configuration reduces electromagnetic convection in the metal bath. The temperature of the liquid steel is controlled by a bichromatic pyrometer; during an experiment, the temperature of the bath is maintained constant, usually at a value between 1600 and 1650 °C.

The gas injection device consists of an alumina tube (7 mm outside diameter, 4 mm inside diameter, 300 mm length), fed with gas through a stainless steel tube, which is connected, outside the furnace, to a mass flow controller and the argon cylinder. The bubbles form at the mouth of an alumina capillary inserted into the injection tube. In order to change the bubble size, we use three different sizes of alumina capillaries (outer diameter: 0.5, 1.2 or 3 mm). Moreover, for a given capillary, the gas flowrate can be modified (between 1 cm$^3$ min$^{-1}$ and 15 cm$^3$ min$^{-1}$) as well as the pressure drop, which enables us to vary the bubble size in a wide range, between 4 and 13 mm (all bubble sizes indicated in the present paper are equivalent volume diameters).

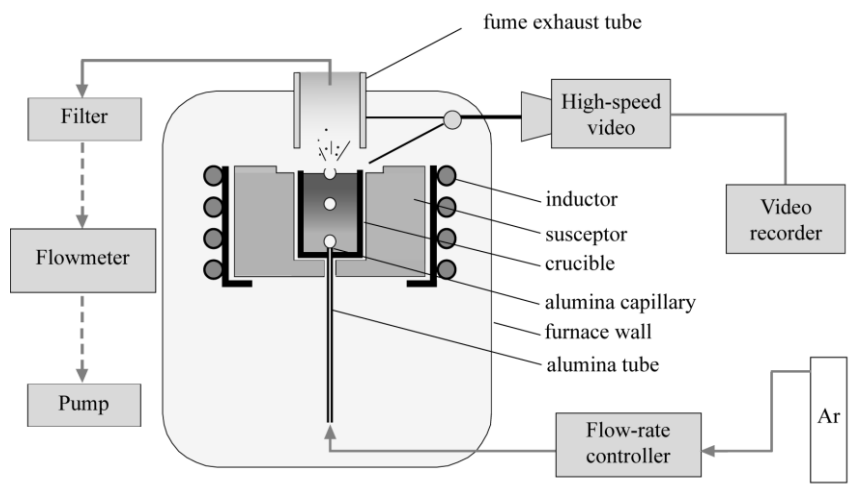

*Figure 13. Schematic representation of the experimental device*

The bursting of bubbles at the surface and the formation of the jet drops are observed by means of a high-speed video camera (Kodak Motion Corder) which makes it possible to film the bath surface at a rate up to 10 000 frames per second. Actually, good-quality images could not be obtained at such a rate because increasing the shooting frequency entails a reduction in image resolution. We therefore selected rates of 5000 frames $s^{-1}$ to record the film break and 1000 frames $s^{-1}$ to observe the formation of the jet drops and to determine the frequency of emergence of the gas bubbles at the surface. The latter frequency is equal to the frequency of the bubble formation at the capillary mouth and thus permits to calculate the bubble size knowing the gas flowrate.

In order to study the film drops, the aerosol formed is exhausted through a rack-mounted tube. The airborne particles are collected on filters inserted in an in-line stainless steel filter holder connected to a flowmeter and a vacuum pump. Two types of filters were used: glass fiber filter (Millipore) for the gravimetric analysis of the particles and PVC membranes (Millipore) for the granulometric analysis and the SEM observation. In order to prevent the saturation of the filters, the exhaust period is limited to 15 seconds for one filter. The flowrate is $4.39 \times 10^{-4}$ $Nm^3$ $s^{-1}$, which corresponds to a gas velocity of 0.4 m $s^{-1}$ at 500 K (typical gas temperature above the bath). According to the Stokes law, it enables to carry particles up to 60 µm in diameter, a size which is larger than that of most of particles contained in the EAF dust. Almost all the particles collected on the filters can be regarded as coming from the steel bath. Indeed, it is possible to remove most of the parasitic particles present in the atmosphere of the furnace by sweeping it with filtered gas. At the beginning of each experiment, the furnace is pumped out and then fed with filtered argon. After one hour of sweeping, there remains in the furnace less than 100 particles with diameters larger than 0.3 µm for 28.3 L of gas and none of these particles have a diameter above 1 µm. Thanks to these experimental precautions, it is possible to obtain a sufficient cleanness of the furnace in order to ensure an accurate determination of the emissions coming from the steel bath.

**3. Results**

3.1. Bubble bursting mechanisms

The analysis of the video sequences reveals that the mechanisms involved in bubble bursting on the free surface of liquid steel are similar to those occurring with air-water systems. Note that no film drops could be observed on the video sequences, due to their smallness and the limited image resolution (one pixel corresponds to 180-200 µm). Nevertheless, their existence is confirmed by the SEM observation of the particles collected on the filters. Two types of sampling were made: with bubbling and without bubbling (i.e. bubbling is turned off for a while). The samples with bubbling show the presence of sphere-like particles (figure 14), with diameters from 0.5 to 40 µm, which do not appear in the samples collected without bubbling. Spherical particles in this size range are of the type, predominant in EAF dust, attributed , in the literature (see part I), to liquid droplet projections. In the present case, these particles are solidified film drops.

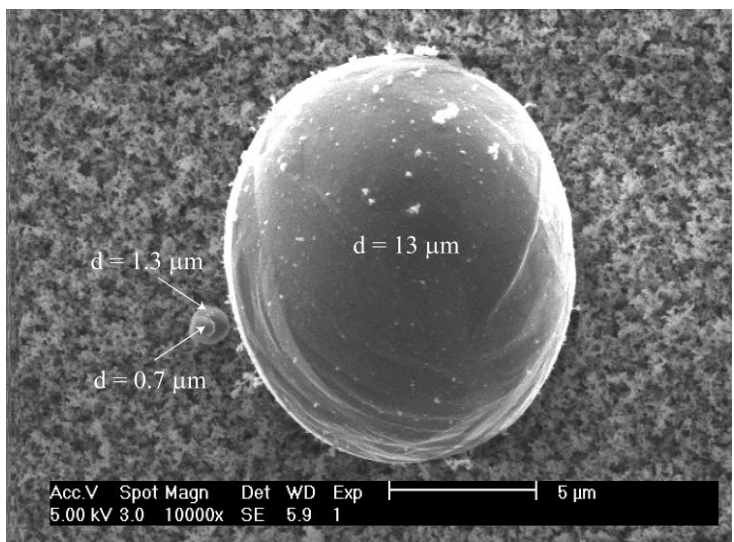

*Figure 14. Several film drops collected in the experimental device*

Unlike the film drops, the projection of jet drops can be observed on the video frames. Figure 15 shows the formation of an upward liquid jet and the projection of one jet drop after a bubble burst at the surface of the liquid steel.

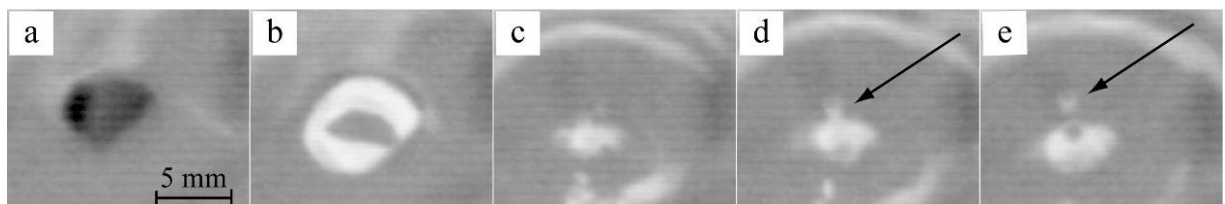

*Figure 15. Frames taken from a video sequence at 1000 frames per second (a: bubble emerging at the bath surface; b: disruption of the bubble cap; c, d: formation of an upward liquid jet; e: emission of a jet drop)*

3.2. Characterization of jet drops

The analysis of video frames is a reliable means to determine precisely the number of ejected jet drops. The results, reported in figure 16, are consistent with those presented in the literature: the probability of jet drop formation, and thus the number of jet drops per bubble, increases as the size of the parent bubble decreases.

By analogy with the correlations proposed by Blanchard [17] and Wu [16] in the case of air-water systems, we derived an exponential law giving the number of ejected drops per bubble ($N_{jet}$) as a function of the bubble diameter ($d_B$, expressed in mm):

$$N_{jet} = 43.4 \, exp(-0.58 \, d_B)$$

The coefficients were calculated by regression from the experimental results (figure 16).

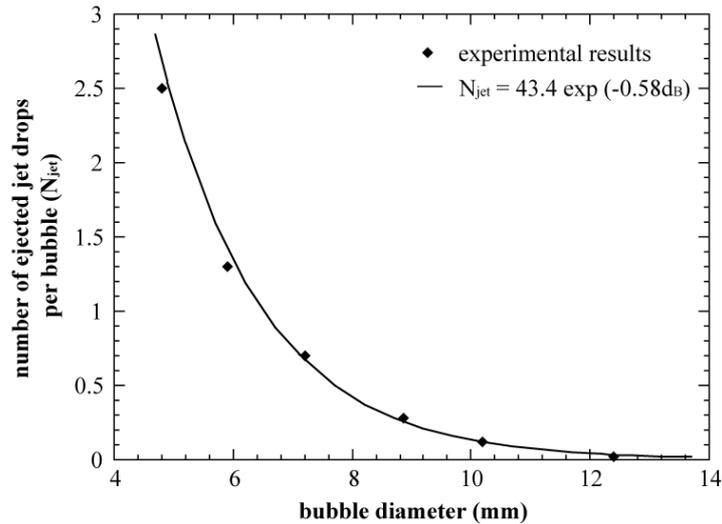

*Figure 16. Number of jet drops versus bubble size*

If the number of jet drops can be determined on the basis of the video sequences, it is much more difficult to measure their sizes because of the poor resolution of the images. Nevertheless, these images show that the jet drops observed are not exhausted by the fume extraction device and fall back into the bath or around the crucible. In order to determine their sizes, the particles gathered around the crucible at the end of each experiment are collected and weighed. These particles are dense metal spheres. From this method, we obtain the size distribution of the jet drops projected during the experiment. The median diameter of each distribution is used in order to characterize the jet drop populations. The results obtained for bubble sizes ranging from 5.5 to 10 mm are reported in figure 17; the jet drop size is proportionnal to the size of the parent bubbles (between 12 and 18 % of the bubble diameter).

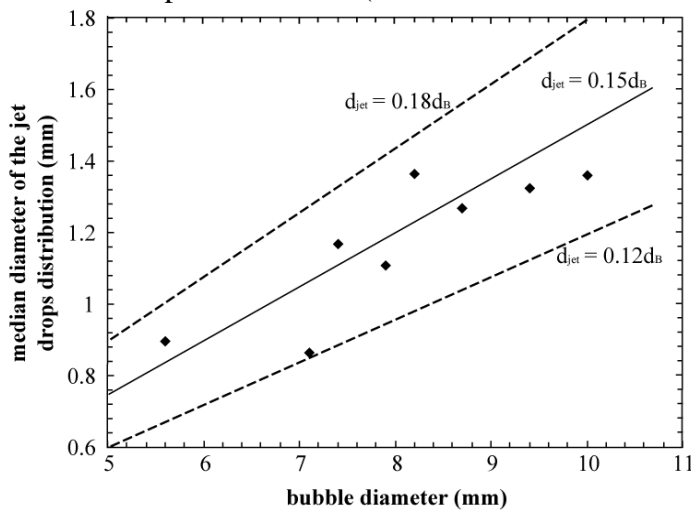

*Figure 17. Size of jet drops versus bubble size*

3.3. Characterization of film drops

To quantify the film drop emission, we analysed aerosols collected during experiments with and without bubbling, using granulometric and gravimetric techniques. Without bubbling, the particles detected come from the vaporization of the steel charge which is clearly visible in the form of fumes inside the furnace.

Granulometric analyses of the experimental dust samples were performed by wet laser diffraction. The Coulter LS 130 granulometer used gives the volume distribution of a suspension of particles ranging from 0.04 µm to $2.10^3$ µm. Dust collected on membranes is dispersed in pure ethyl alcohol and desagglomerated by ultrasonic and mechanic agitation following the procedure described in figure 18. Such a procedure leads to an optimal desagglomeration of the particles and a good reproducibility of the measurements.

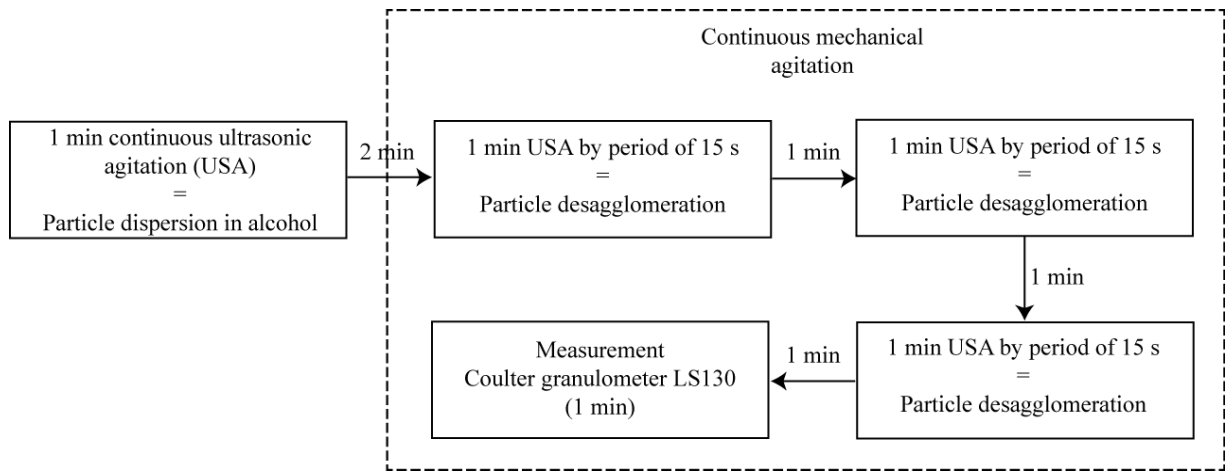

*Figure 18. Procedure for the preparation of the particle suspension and for the granulometric analysis*

Figure 19 shows the typical results of a granulometric analysis of two kinds of dust samples (with and without bubbling). They confirm the existence of two particle populations: submicronic particles coming from the vaporization of the liquid steel, and film drops. Concerning the first population, we do not have yet a satisfactory explanation for the presence of three modes. The important feature in the figure is the apparition of the biggest particles in the sample collected with bubbling.

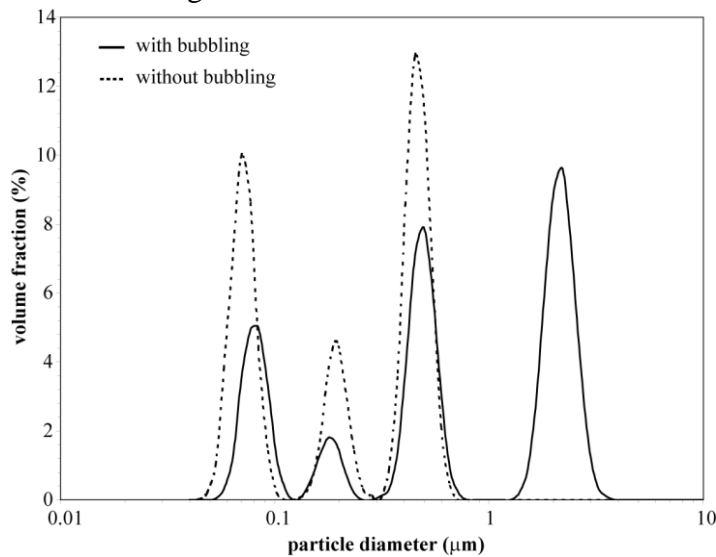

*Figure 19. Size distribution of two dust samples collected with and without bubble bursting (bubble diameter: 7 mm)*

From such granulometric results and knowing the amount of dust collected with and without bubbling (see below), it is possible to obtain the size distribution of film drops by "substracting" both types of results. Figure 20 presents the distributions obtained for three bubble sizes. It can be seen that the size spectrum broadens up when the bubble size increases, even if most of the film drops remain under 20 µm, as in EAF dust samples. This phenomenon is confirmed by the SEM observation of the samples.

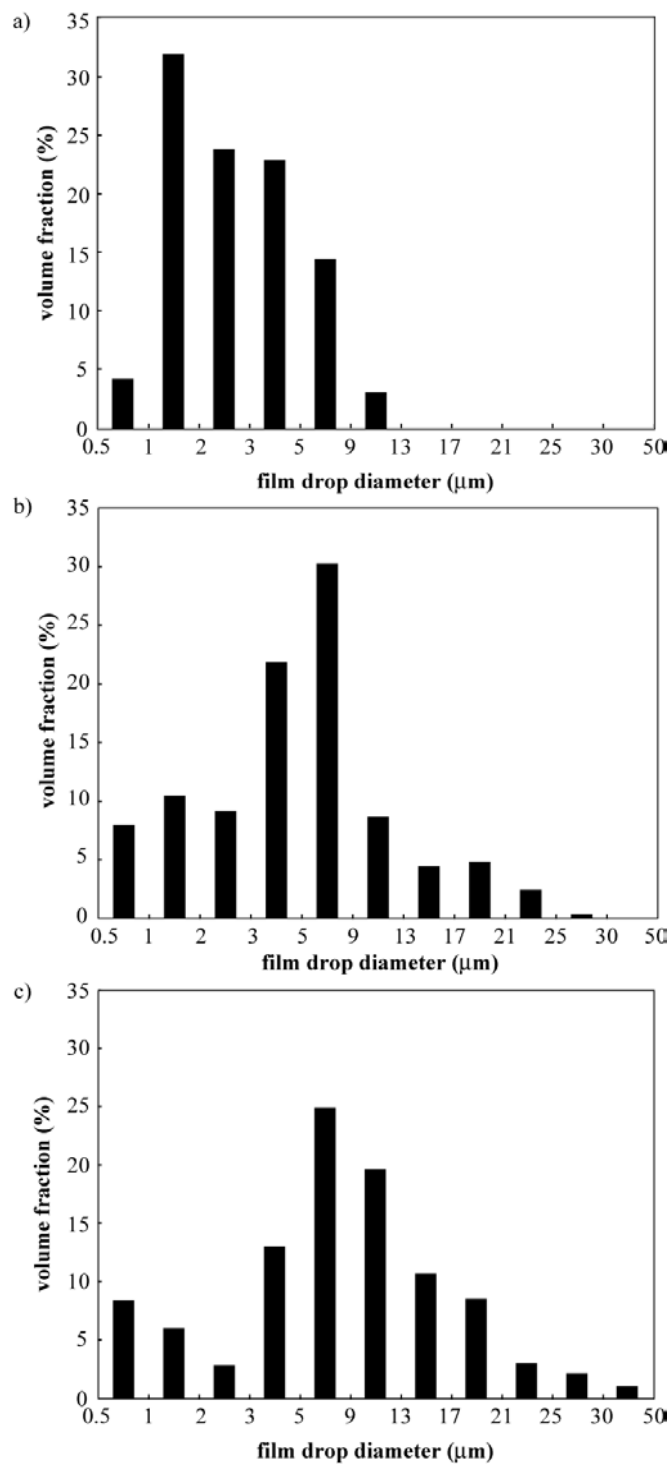

*Figure 20. Size distribution of film drops produced by bubble bursting.*
*Bubble diameter: a) 8.3 mm, b) 10.2 mm, c) 12.5 mm*

The mass of particles collected was determined by weighing of the glass fiber filters before and after each experiment. The difference between the results of the experiments with- and without bubbling gives the mass of film drops collected. The results are gathered in table 1; for an easier and more meaningful comparison, the amounts have been referred to one bubble burst ($M_B$ in table 1) and to the volume of injected gas ($M_G$ in table 1).

Table 1. Mass of film drops as a function of the parent bubble size

| $d_B$ (bubble size) | $M_B$ (mass of projections for one bubble burst) | $M_G$ (mass of projections for 1 m$^3$ of injected gas) |
|---|---|---|
| 4 mm | not detected | < 1 g/m$^3$ |
| 4.6 mm | not detected | < 1 g/m$^3$ |
| 7.2 mm | 3.75 µg/bubble | 19.2 g/m$^3$ |
| 9.1 mm | 10.1 µg/bubble | 25.6 g/m$^3$ |
| 10.7 mm | 18.9 µg/bubble | 29.5 g/m$^3$ |
| 12.6 mm | 30.3 µg/bubble | 29 g/m$^3$ |

The results clearly show that the amount of film drops produced by one bubble burst greatly increases with the bubble size. For the smallest bubbles, under 4.6 mm in diameter, no film drop was detected. More precisely, the amount of film drops was so low that it could not be detected, the difference in weight between the samples with- and without bubbling being below the balance sensitivity ($10^{-5}$ g). The mass of film drops for 1 m$^3$ of injected gas also increases with the bubble size as long as this bubble size is lower than 11 mm. Above this size, we can notice a plateau or a slight decrease in the amount of film drops. It would be interesting to investigate further this phenomenon, as well as to relate the amount of film drops to the mass of the bubble cap prior to bursting.

## 4. Interpretation

In EAF, 60 % of dust come from projections of liquid metal and slag (see part I). These projections from the bath can be attributed to the CO-bubble burst. The mechanisms involved in the formation of those projections in EAF can be slightly different from those met in our experiment, particularly owing to the presence of a slag, either foaming or not. Nevertheless, the results we have obtained give useful information for the understanding and the quantification of dust formation in EAF.

**Jet drops** come from the disintegration of the upward jet created after the removal of the bubble cap. Their number increases when the bubble size decreases, and their size represents 12 to 18 % of the parent bubble size. The size of CO-bubbles formed in EAF remains little known. However, analyses of foaming slag samples and numerical calculations indicate that their sizes are probably between 2 and 20 mm [8]. According to our results, such bubbles are expected to produce jet drops whose sizes vary from 0.2 mm to almost 4 mm, which is much larger than most of the particles found in EAF dust samples. As observed in our laboratory furnace, jet drops are not exhausted by the fume extraction system and are likely to fall back into the steel bath. Jet drops can thus hardly contribute to dust formation from bubble burst in EAF.

**Film drops** are emitted during the disintegration of the liquid cap which covers the bubble at the surface of the bath. Their morphology and their size range are very close to those of the particles contained in EAF dust. The amount of projections produced by bubble burst in EAF varies between 0.016 and 0.028 kg m$^{-3}$ [8]. These figures are close to those derived from our laboratory experiments (see table 1). Further associated with the conclusion of the jet drop size study, they show that, when a bubble bursts, it is mostly the film drops that contribute to dust formation. Our results also reveal a significant decrease of the amount of film drops resulting from bubble burst when the bubble size decreases. Moreover we brought out the existence of a critical bubble size (around 4.5 mm) under which no film drop is detected. While this phenomenon was already evidenced by Spiel [14] for air-water systems, it had never been observed in the case of liquid steel.

From these results, it appears that it should be possible to reduce dramatically the amount of dust produced in EAF by decreasing the CO-bubble sizes, ideally between 1 and 4 mm. The latter bound prevents the film drop formation and the former one avoids the emission of jet drops small enough to be carried up. Such an objective may be difficult to reach since the CO-bubble formation is a rather spontaneous process. Nevertheless, a solution could be to better control the decarburization reaction, for example by favoring nucleation at the expense of growth.

## CONCLUSION

From the study of the morphological and mineralogical characteristics of EAF dust samples combined to the knowledge of the EAF steelmaking process, we could determine the mechanisms of EAF dust formation. Among the different sources of emission, the major one is the projection of liquid droplets by bubble burst at the liquid bath surface. We therefore designed an experimental device to observe the gas bubble burst at the surface of liquid steel by means of high-speed video, and to quantify the resulting projections by granulometric and gravimetric analyses. The phenomena involved are similar to those taking place in the case of an air bubble bursting at water surface and result in the emission of two types of droplets: film drops and jet drops. Only film drops take part to the formation of EAF dust, jet drops are too big to be exhausted and fall back into the liquid bath. We have also shown that the amount of film drops decreases with the size of the parent bubble. When the bubble size reaches 4.5 mm, no film drop is emitted. The bubble size is therefore a key parameter for the reduction of dust produced in EAF.

The continuation of the present piece of work will consist in studying the influence of a slag at the surface of the bath, as well as that of surfactants, on the bubble-burst process.